\documentclass[review]{elsarticle}

\usepackage{amssymb}

\usepackage{amsthm}

\usepackage{amsmath}

\usepackage[a4paper,bindingoffset=0.2in,left=1in,right=1in,top=1in,bottom=1in,footskip=.25in]{geometry}

\usepackage{color,soul}

\usepackage{booktabs}

\usepackage{rotating}

\makeatletter
\def\ps@pprintTitle{%
 \let\@oddhead\@empty
 \let\@evenhead\@empty
 \def\@oddfoot{}%
 \let\@evenfoot\@oddfoot}
\makeatother

\journal{Journal of Membrane Science}

\begin{document}
\biboptions{sort&compress}

\abovedisplayshortskip=-13pt
\belowdisplayshortskip=4pt
\abovedisplayskip=-13pt
\belowdisplayskip=4pt

\begin{frontmatter}



\title{Nernst-Planck transport theory for (reverse) electrodialysis: \\ II. Effect of water transport through ion-exchange membranes}

\author{M. Tedesco}
\author{H.V.M. Hamelers}
\author{and P.M. Biesheuvel \corref{cor1}} \cortext[cor1]{\emph{Corresponding Author} (P. M. Biesheuvel): maarten.biesheuvel@wetsus.nl}

\address{Wetsus, European Centre of Excellence for Sustainable Water Technology, \\ Oostergoweg 9, 8911 MA Leeuwarden, The Netherlands}

\begin{abstract}
Transport of water through ion-exchange membranes is of importance both for electrodialysis (ED) and reverse electrodialysis (RED). In this work, we extend our previous theory [J. Membrane Sci., \textbf{510}, (2016) 370-381] and include water transport in a two-dimensional model for (R)ED. Following a Maxwell--Stefan (MS) approach, ions in the membrane have friction with the water, pore walls, and one another. We show that when ion-ion friction is neglected, the MS--approach is equivalent to the hydrodynamic theory of hindered transport, for instance applied to nanofiltration. After validation against experimental data from literature for ED and RED, the model is used to analyze single-pass seawater ED, and RED with highly concentrated solutions. In the model, fluxes and velocities of water and ions in the membranes are self-consistently calculated as function of the driving forces. We investigate the influence of water and coion leakage under different operational conditions. 
\vspace{.3cm}
\end{abstract}


\end{frontmatter}

\section{Introduction}
\label{Intro}

\indent In electromembrane processes, ion-exchange membranes (IEMs) are used to desalinate water, to selectively remove ions or other charged molecules, and to generate energy from salinity differences \citep{Tedesco2016}. IEMs are membranes that contain fixed charges and allow counterions to pass, while in the ideal case coions and water are completely rejected. For all applications of electromembrane processes, a general membrane transport theory describes ion and water flow as function of gradients in electrical potential, salt concentration, and pressure in a self-consistent manner without ad-hoc assumptions. 

\indent The complexity of such a general theory depends on the behavior of the membrane: in the ideal case, only counterions pass through the membranes, and the mathematical description of membrane transport is the least complex \citep{Sonin1968}. However, in most cases membranes cannot be described as ideally selective barriers, as they are partially crossed by coions and water, resulting in a decrease of process performance. This non-ideal behavior of membranes requires the simultaneous description of transport of counterions, coions, and water. Co-ion transport was discussed in detail in our previous work, but we did not yet discuss water transport~{\citep{Tedesco2016}}.

\indent Water transport through ion-exchange membranes has been addressed in the literature both experimentally \citep{evans2006,Izquierdo2012,han2015} and theoretically  \citep{Kedem1961,Holt1981,jiang2015}. Four mechanisms are often considered to cause water transport: I. A hydrostatic pressure difference across the membrane; II. Osmosis, which is water transport due to an osmotic pressure difference across the membrane; III. Electro-osmosis, water transport due to ion-water friction; and IV. Water transported in the hydration shell around the ions, where water molecules are tightly held. In the present work, we focus on mechanisms I-III and neglect mechanism IV. Note that we will use the words ``water'', ``solvent'', and ``fluid'' interchangeably.

\indent Although different theoretical approaches have already been proposed, combining ion and water transport in a general model which describes both Electrodialysis (ED) and Reverse Electrodialysis (RED) is still challenging. As an example of such challenges, transport of water is affected by differences across the membrane in salt concentration, electric potential and hydrostatic pressure, as well as by friction with ions and pore walls, and all effects must be included simultaneously~\cite{Peters2016}.

\begin{figure}[ht]
\centering
\includegraphics[width=\textwidth]{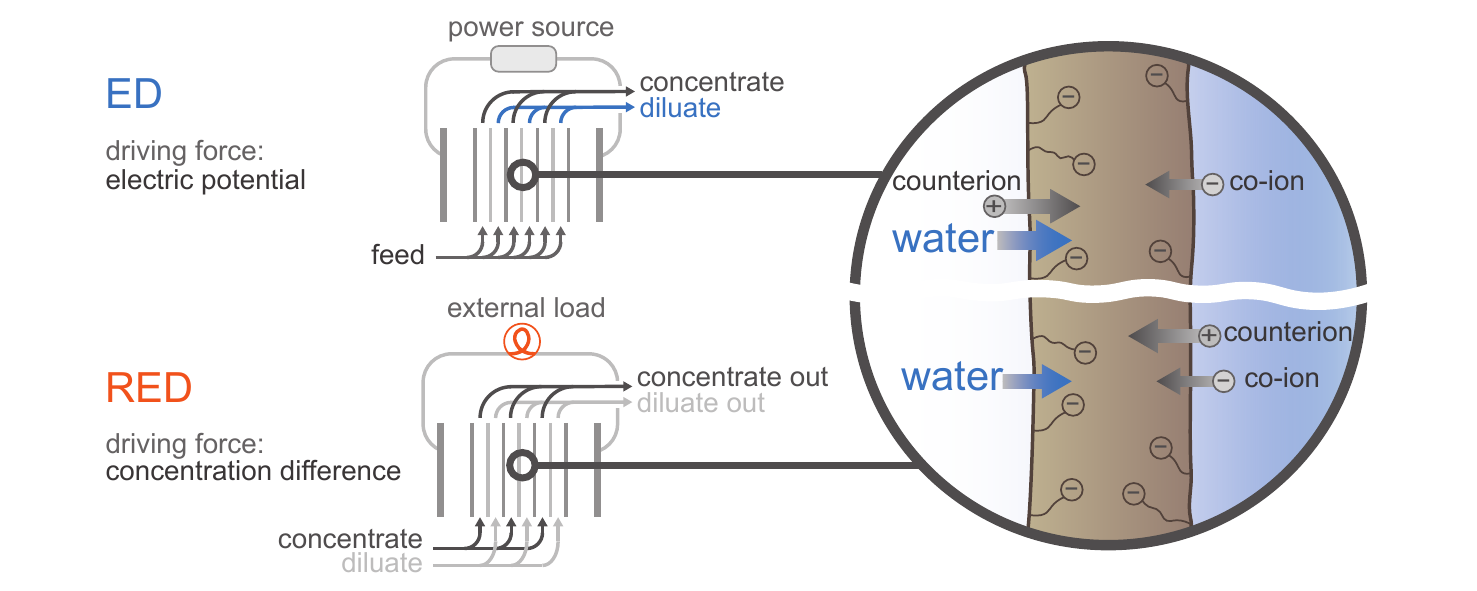}
\caption{Principle of electrodialysis (ED) and reverse electrodialysis (RED). The ED process is used as a separation process of aqueous streams by applying a voltage across membranes, while RED harvests the concentration difference between two salt solutions for power production. The repeating unit of the system is the cell pair which consists of a concentrate channel, a cation exchange membrane (CEM), a diluate channel, and an anion exchange membrane (AEM).}
\label{fig:FIG_principle}
\end{figure}

\indent Previously, we developed a two-dimensional electromembrane model based on the approach by Sonin and Probstein \citep{Sonin1968}, focusing on the effect of coion transport in ED and RED \citep{Tedesco2016} (See Fig. \ref{fig:FIG_principle} for a brief description of these processes). Our work showed that in RED coion transport reduces the power density by up to 20\%, while for ED the energy consumption increases by at least a factor of three compared to the ideal case \cite{Tedesco2016}. In the present work, we extend the model by including fluid (water) transport through the membranes. We focus on the modeling scale of a single cell pair, because most of the relevant transport phenomena of the (R)ED process relate to the cell level. Stack-level modeling is of importance to identify hydraulic leakage between cells \citep{Tanaka2004pressure}, non-uniform flow distribution \citep{Kodym2012}, and parasitic currents in the manifolds \citep{Veerman2008parasitic}, but will not be discussed here. We validate the model against experimental data from literature (both for ED and RED), and use the model to describe quantitatively the effects of coion and water transport.

\section{General aspects of ion transport in liquid-filled membrane pores}
\label{sec:Theory}

\indent In (reverse) electrodialysis, transport phenomena in the spacer channel have often been the primary focus of theoretical studies \citep{Teorell1953,Probstein1972,Lacey1980, Pintauro1984,Guzman-Garcia1990,Higa1990, Fila2003,Volgin2005,Fidaleo2005,Veerman2011, Tanaka2015},  with membrane transport described by a constant permselectivity (set to 100\%, in many cases) and water flow neglected (with \citep{Fila2003} a notable exception). 
Instead, in the present work we combine a two-dimensional (2D) model for the spacer channels with a detailed description of transport of ions and water through the membranes \citep{mehta1976,Meares1981}, thus providing a full description of transport phenomena in the channels and in the membranes. In the 2D model, one coordinate axis runs through the cell (from entrance to exit), and one coordinate is directed towards and through the membranes, see Fig. \ref{fig:FIG_scheme}. 

For the modeling of ion diffusion and electromigration in the spacer channels, we assume that the ions have friction only with the fluid (ion-solvent, or ion-fluid, interaction), while in the membrane we consider that ions also have friction with the membrane matrix (`ion-wall' or `ion-matrix' interaction), as well as with other ions (`ion-ion' interaction), as described by Maxwell-Stefan (MS) theories \citep{mehta1976, Kraaijeveld1995, Amundson2003,delacourt2008}. Following the Maxwell-Stefan framework, the total force acting on an ion $i$ in the membrane can be described as 

\begin{equation}
	- \nabla \tilde{\mu}_i = R_\text{g}\: T\: \sum_j{f_{i-j} \left( \textbf{v}_i - \textbf{v}_j \right)}
\label{eq:MS0}
\end{equation}

\noindent where $\tilde{\mu}_i$ is the electrochemical potential of ion $i$, $R_\text{g}$ the gas constant, $T$ temperature, $f_{i-j}$ a friction factor between ion $i$ and phase $j$ (which can be the water, the membrane matrix, or another ion), while $\textbf{v}_i$ and $\textbf{v}_j$ are velocities of $i$ and $j$. 

\indent In this work, we assume ideal thermodynamics with ions as point charges and consider the water as a continuum fluid. For ideal thermodynamic behavior of the ions, $\tilde{\mu}_i$ is then equal to

\begin{equation}
\tilde{\mu}_i=\tilde{\mu}_{i,0}+R_\text{g} T\ln c_i+RT z_i\phi
\end{equation}

\noindent where $c_i$ is ion concentration, $z_{i}$ ion valence, $\phi$ the dimensionless electric potential (which can be multiplied by $V_T=RT/F$ to obtain a dimensional voltage), after which Eq. (\ref{eq:MS0}) can be written in a single direction, $x$, as 

\begin{equation}
	-  \frac{\partial{\ln c_i}}{\partial{x}} - z_i\: \frac{\partial{\phi}}{\partial{x}}  = \frac{1}{D_{i-F}} \left( v_i - v_{F} \right) +  f_{i-m} \: \left( v_i - v_m \right) + \beta c_k \left( v_i - v_k \right)
\label{eq:MS2}
\end{equation}
where $D_{i-F}$ is an ion-fluid diffusion coefficient, $D_{i-F}=1/f_{i-F}$, $f_{i-m}$ is a friction factor between ion $i$ and the membrane matrix (which has zero velocity, $v_m=0$), and $v_{F}$ is the fluid velocity in the membrane. The last term in Eq.~(\ref{eq:MS2}) describes friction of ion $i$ with another ion, $k$, and includes a linear dependence on the concentration of the other ion, and on the velocity difference. Note that in this work concentrations are defined per unit open volume, i.e., per unit aqueous volume in the membrane (or, in spacer channel) \cite{Galama2013}, while velocities and fluxes are defined per unit total area of a channel or membrane. 

\indent Neglecting ion-ion friction (setting $\beta$ to zero), Eq. (\ref{eq:MS2}) can be rewritten by introducing a modified diffusion coefficient, $D^*_i$, which is defined as 

\begin{equation}
	\frac{1}{D^*_i} = \frac{1}{D_{i-F}} + f_{i-m}
\label{eq:D_star}
\end{equation}

\noindent resulting for the ion velocity in the membrane in~\citep{Thibault2015}

\begin{equation}
	v_i = \frac{D^*_i}{D_{i-F}} \: v_{F} - D^*_i \left( \frac{\partial{\ln c_i}}{\partial{x}} + z_i\: \frac{\partial{\phi}}{\partial{x}} \right) .
\label{eq:MS4}
\end{equation}

\noindent Eq. (\ref{eq:MS4}) gives the velocity of ion $i$ in an ion-exchange membrane, taking into account ion friction with the fluid and with the membrane matrix. Interestingly, Eq. (\ref{eq:MS4}) is formally equivalent to an expression from hydrodynamic theory for hindered transport, in which filtration of particles through a pore is described by means of two `hindrance' factors, $K_{c}$ and $K_{d}$, where ``$c$'' and ``$d$'' refer  to ``convection'' and ``diffusion'', respectively \citep{Deen1987,Thibault2015}. We rewrite Eq. (\ref{eq:MS4}) to

\begin{equation}
	v_i = K_{c,i} \: v_{F} - K_{d,i}\:\frac{\epsilon}{\tau}\: D_{i,\infty} \left( \frac{\partial{\ln c_i}}{\partial{x}} + z_i\: \frac{\partial{\phi}}{\partial{x}} \right)
\label{eq:Deen}
\end{equation}
\noindent where the equalities $K_{c,i}=D^*_i/D_{i-F}$ and $K_{d,i}=D^*_i/\frac{\epsilon}{\tau} D_{i,\infty}$ connect Eqs. (\ref{eq:MS4}) and Eq. (\ref{eq:Deen}), and where $D_{i,\infty}$ is the ion diffusion coefficient in free solution, $\epsilon$ membrane porosity, and $\tau$ pore tortuosity. The hindrance factors, $K_{c}$ and $K_{d}$, depend on hydrodynamic conditions in the liquid-filled pore, and a number of correlations have been reported to calculate $K_{c}$ and $K_{d}$ for the case of neutral (spherical) molecules in straight cylindrical pores \citep{Brenner1977, Noordman2002}. 

\indent Using Eq. (\ref{eq:Deen}), the ion flux through the membrane, $J_i=c_i \:v_i$, is now given by \citep{Deen1987, Lanteri2009, Szymczyk2010} 

\begin{equation}
	J_i = K_{c,i} \: c_i \: v_{F} - K_{d,i}\:\frac{\epsilon}{\tau}\: D_{i,\infty} \left( \frac{\partial{c_i}}{\partial{x}} + z_i\: c_i\: \frac{\partial{\phi}}{\partial{x}} \right)
\label{eq:J_i}
\end{equation}

\noindent which is a general expression for ion transport including ion-fluid and ion-membrane friction. Note that if ion-membrane friction is neglected ($f_{i-m}=0$), that $D_i^*=D_{i-F}$, $K_{c,i}=1$, and Eq. (\ref{eq:J_i}) simplifies to

\begin{equation}
	J_i = c_i \: v_{F} - D_{i-F} \left( \frac{\partial{c_i}}{\partial{x}} + z_i\: c_i\: \frac{\partial{\phi}}{\partial{x}} \right)
\label{eq:NP_x}
\end{equation}

\noindent which is the well-known extended Nernst-Planck (NP) equation~\cite{Higa1990,Fila2003,Volgin2005,Fidaleo2005, Biesheuvel2011b}. Thus, Eq. (\ref{eq:J_i}) is a modified NP-equation that includes also ion-wall friction, where $K_d$ and $K_c$ have values between 0 and 1.

\indent The results obtained above demonstrate that, neglecting ion-ion interactions, Maxwell--Stefan theory gives a framework equivalent to the hydrodynamic theory for hindered transport of solutes through narrow pores \citep{Deen1987}. This equivalence allow us to implement information about solute hindrance into the MS framework. This is advantageous because the hydrodynamic models for hindered transport do not include a description of fluid flow, but this can be included in the MS framework \citep{Deen1987}.


\section{Model development}
\label{sec:ModelDevelopment}

\begin{figure}[ht]
\centering
\includegraphics[width=0.70\textwidth]{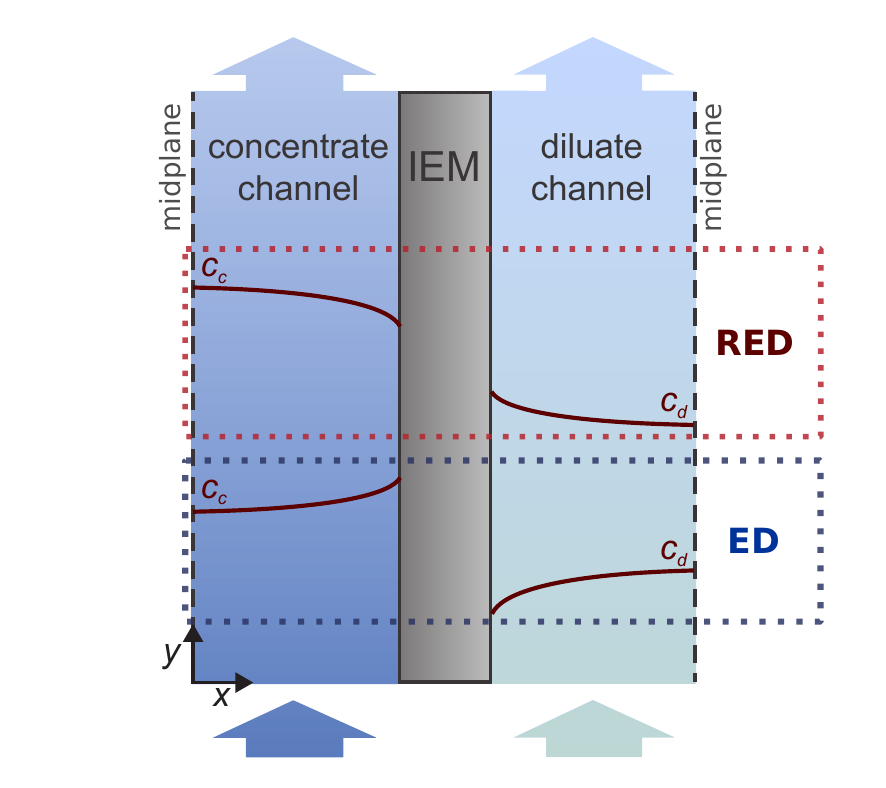}
\caption{Schematic diagram of the model geometry. The computational domain consists of half of a concentrate channel, one ion exchange membrane (IEM) and half of a diluate channel. Symmetry in $x$-direction is assumed around the midplanes. Typical concentrations profiles for ED and RED are sketched.}
\label{fig:FIG_scheme}
\end{figure}

\indent Having laid out the general membrane ion transport model, we show how it can be incorporated in a 2D cell model. In the 2D model, ion concentrations in the spacer channel vary in two directions: from membrane to membrane (the $x$-direction), and from entrance to exit of the cell (along the membrane, $y$-direction). A co-current flow arrangement is considered, where both feed streams flow in the same $y$-direction, while the electric field acts in $x$-direction, perpendicular to flow, see Fig. \ref{fig:FIG_scheme}. We assume steady-state operation, and assume that both in the channel and membranes all diffusive and electromigration fluxes (thus current density) run exactly in $x$-direction. We impose that the cell pair voltage is the same at each {$y$}-coordinate, and allow the current density to change with {$y$}. In the $y$-direction in the channels, we assume absence of diffusion and electromigration, and thus in this direction the ions are only advected with the fluid. For the fluid, we assume a plug flow profile ($y$-component of the fluid velocity, $v_y$, independent of $x$-coordinate). Because of fluid transport through the membrane, $v_y$ will be $y$-dependent. Note that, for sufficiently thin channels (as for RED), the exact flow profile (parabolic or plug flow) has a negligible effect on the salt flux through the membranes {\citep{Tedesco2016}}.

\indent The modeling framework presented here can be used to describe a complete cell pair, i.e., the repeating unit of an (R)ED system. However, to simplify the calculation, we assume in the present work that both the salt molecule and the membranes have ``symmetric'' properties. In particular, we consider that both AEMs and CEMs have the same physical properties, like porosity, thickness, and magnitude of the membrane charge, and that counterions in the membranes (anions in AEMs, and cations in CEMs) have the same behavior in terms of ion-fluid and ion-membrane friction. The same is assumed to hold for coions as well. In the spacer channel, we assume that the diffusion coefficient of cation and anion is the same (this assumption is not made for the membrane). Because of these assumptions, the ``repeating unit'' in the calculation can be reduced to only one membrane and two ``half''-channels, see Fig. \ref{fig:FIG_scheme}~\cite{Tedesco2016,Sonin1968}.

\subsection{Modeling the spacer channel}
\label{sec:ModelingTheSpacerChannel}

\indent In the 2D model, the ion mass balance in the flow compartment (spacer channel) is given by

\begin{equation}
	\epsilon \: \frac{\partial{c_{i}}}{\partial{t}} + \nabla{\cdot J_{i}} = 0
	\label{eq:mass_balance}
\end{equation}

\noindent where $J_{i}$ is the ion flux (defined per total cross-sectional area), and $\epsilon$ is the channel porosity (volume left open by spacer mesh). The ion flux, $J_{i}$, can be described by the extended Nernst-Planck (NP) equation
\begin{equation}
	J_{i}=c_{i}\: \textbf{v} - D_{i,\text{d}} \: \nabla{c_{i}} - D_{i,\text{e}} z_{i} \: c_{i} \nabla{\phi}
	\label{eq:NP}
\end{equation}

\noindent where $\textbf{v}$ is the fluid flow velocity in the channel (also defined per total cross-sectional area, i.e., the ``empty tube'' velocity), and $D_{i,\text{d}}$ and $D_{i,\text{e}}$ are effective ion diffusion coefficients for diffusion and electromigration. For electromigration, $D_{i,\text{e}}=\epsilon/\tau \cdot D_{i.\infty}$ where $D_{i,\infty}$ is the ion diffusion coefficient in free solution, $\epsilon$ is the spacer open fraction (porosity) and $\tau$ is the tortuosity factor in the channel. For the diffusional term, the effective coefficient is larger because dispersion effects play a role: because the spacer mesh creates tortuous pathways, the fluid mixes, and thus the concentration is more leveled out. This can be described by adding a dispersion term to the flux equation, which can be combined with the diffusive term and leads to an increase in the effective coefficient for diffusion (and not for electromigration); thus $D_{i,\text{d}} > D_{i,\text{e}}$. Substituting Eq. (\ref{eq:NP}) into Eq. (\ref{eq:mass_balance}) leads to

\begin{equation}
	\epsilon \: \frac{\partial{c_{i}}}{\partial{t}} =  D_{i,\text{d}}  \: \nabla^2 c_{i} + D_{i,\text{e}} \: z_{i} \: \nabla \cdot \left( c_{i} \nabla{\phi} \right)  - \nabla \cdot \left({c_{i} \textbf{v}} \right)
	\label{eq:mass_bal_NP}
\end{equation}

\noindent which applies to both ions in the spacer channels. In the present work, we consider the system in steady--state, and thus $\partial{c_i}/\partial{t}=0$, while both streams are assumed to contain only a completely dissociated 1:1 salt. We assume that in the spacer channels cations and anions have the same diffusion coefficient ($D_{+,\text{d/e}}=D_{-,\text{d/e}}=D_\text{d/e}$), and we neglect diffusion and electromigration in the axial \mbox{($y$-)direction} (an assumption discussed in detail in ref. \citep{Sonin1968}). According to the electroneutrality condition, the concentrations of cation and anion at each point in the spacer channel are equal to each other, thus $c_{+}=c_{-}=c$. Therefore, adding up Eq. (\ref{eq:mass_bal_NP}) for anions and cations results in the salt balance
\begin{equation}
\frac{\partial}{\partial{x}} \left(c \> v_{x} - D_\text{d}\: \frac{\partial{c}}{\partial{x}} \right) + \frac{\partial{}}{\partial{y}}\left( c \> v_y \right) = 0.
	\label{eq:mass_bal_NP2}
\end{equation}
 
\indent In order to keep $v_y$ independent of $x$ at all $y$-positions (related to the assumption of plug flow), the fluid velocity in $x$-direction, $v_x$, relates to the fluid velocity in the membrane, $v_{F}$, by

\begin{equation}
	\frac{\partial v_y}{\partial y}=-\frac{\partial v_x}{\partial x}=\pm \frac{v_{F}}{h}.
\label{eq:v_y}
\end{equation}

\noindent where $h$ is the channel half-width and the sign ``$+$'' is used for the diluate channel, and ``$-$'' for the concentrate channel (following the geometry outlined in Fig. \ref{fig:FIG_scheme}). The ionic current density across the flow channel (in the $x$-direction) is 
\begin{equation}
	J_\text{ch}=J_{+} - J_{-}	
\label{eq:J_ch}
\end{equation}

\noindent where + and - refer to cation and anion. Note that in Eq. (\ref{eq:J_ch}) $J_\text{ch}$ is expressed in mol/m$^2$/s, and must be multiplied by $F$ to obtain a current density in A/m$^2$, of which the average over the cell will be denoted by $I$. We define two average salt concentrations, $\langle c \rangle $ and $\langle c \rangle ^\dagger$ which are both averages across the width of the channel (in $x$-direction), and which both depend on $y$-position. Evaluated at the exit of the cell, one of them, $\langle c \rangle $, is the ``mixed cup'' effluent concentration. These two concentrations are given by

\begin{equation}\label{eq:average}
\langle c \rangle= \frac{1}{h}\int^h_0 c(x) \text{d}x\hspace{5mm},\hspace{5mm}
\frac{1}{\langle c \rangle ^{\dagger}}=\frac{1}{h} \int^h_0 \frac{1}{c(x)} \text{d}x.
\end{equation}

\indent Combining Eq.~(\ref{eq:J_ch}) with the NP-equation~(\ref{eq:NP}), and taking into account that the current density, $J_\text{ch}$, is not a function of $x$ (only of $y$), we can integrate across the width of the flow channel (in $x$-direction) and arrive at

\begin{equation}
	J_\text{ch}= - \: 2 \: \langle c \rangle ^\dagger \: D_\text{e}  \: \frac{\Delta \phi}{h}
\label{eq:J_ch_avg}
\end{equation}

\noindent where $\Delta \phi$ is the potential difference across half of the channel.  

\indent An overall salt balance over a ``slice'' with thickness d$y$ in the channel relates the ions flux through the membrane to $\langle c \rangle$ according to

\begin{equation}
	J_{\text{ions},m}=J_{+,m}+J_{-,m}=\pm 2h \frac{\text{d} \left( v_y \cdot \langle c \rangle \right)}{\text{d}y}.
\label{eq:J_ions_mem}
\end{equation}

\indent Interestingly, because of the use of Eqs. (\ref{eq:J_ch_avg}) and (\ref{eq:J_ions_mem}), no boundary conditions need to be given at the solution/membrane interface for the gradient in concentration or potential. In the middle of the flow channels, symmetry around the midplane leads to $\partial c / \partial x = 0$.

\subsection{Modeling the membrane}
\label{sec:ModelingMembrane}

\indent In the membrane, local electroneutrality holds at each position,

\begin{equation}
	c_{+} - c_{-} + \omega X= 0
	\label{eq:LEN}
\end{equation}

\noindent where $\omega$ is the sign of the fixed membrane charge ($\omega=+1$ for AEMs, and $\omega=-1$ for CEMs), and $X$ is the molar concentration of membrane charge, defined per unit volume of solution phase in the membrane \cite{Galama2013}. 

\indent The fluid velocity in the membrane, $v_F$, can be calculated from \citep{Biesheuvel2011b, Yaroshchuk2011}

\begin{equation}
	\frac{\partial{P^t}}{\partial{x}} = f_{F-m} \left( v_m - v_F \right) \:  + \sum_i \frac{c_i}{D_{i-F}} \left(v_i - v_F \right)
\label{eq:fluid_2}
\end{equation}

\noindent with $f_{F-m}$ the fluid-membrane friction factor, and $P^t$ the total pressure, which is hydraulic pressure, $P^h$, minus osmotic pressure \citep{Sonin1976selegny, Spruijt2014, Peters2016}. The total pressure is invariant across the solution/membrane boundary (i.e., across the EDL). Note that in Eq. (\ref{eq:fluid_2}) $P^t$ has dimension mol/m$^3$, and can be multiplied by $RT$ to a pressure with unit Pa. In Eq. (\ref{eq:fluid_2}), we assume the ions to be point charges (i.e., without volume), and the summation i runs over all ions. Note that when there are neutral species in the membrane, they must also be considered in Eq. ({\ref{eq:fluid_2}}). For ions as ideal point charges, the osmotic pressure is equal to the total ion concentration (not salt concentration), multiplied by $RT$. Note that Eq. (\ref{eq:fluid_2}) is also valid when the ions have friction with the membrane matrix and/or with one another. Interestingly, if friction between ions and membrane matrix is neglected (and $v_m=0$), Eq. (\ref{eq:fluid_2}) simplifies to 

\begin{equation}
	\frac{\partial{P^h}}{\partial{x}} + f_{F-m}\: v_F = \omega\:X \: \frac{\partial{\phi}}{\partial{x}}
\label{eq:fluid_Sonin}	
\end{equation}

\noindent which is the classical result reported by Sonin \cite{Sonin1976selegny} and Schl\"{o}gl \citep{Schlogl1964}.

\indent Transport of ions in $x$-direction through the membrane is described by Eq.~(\ref{eq:MS2}) evaluated separately for the anion and cation, while considering steady-state. Steady state implies that, at a given $y$-coordinate, the ion flux $J_i = v_i \cdot c_i$ is constant across the membrane, both for anion and cation (also the fluid velocity $v_F$ is constant across the membrane because we neglect the volume of the ions). Instead, ion velocities are not invariant across the membrane, as will be shown in Section~\ref{sec:ResultsAndDiscussion}. Using Eqs. (\ref{eq:J_ch}) and (\ref{eq:J_ions_mem}), cation and anion fluxes in the membrane, $J_+$ and $J_-$, relate to the total ions flux, $J_{\text{ions},m}$, and current density, $J_\text{ch}$. 

\indent To solve for the ion transport in the membrane, Eq. (\ref{eq:MS2}) is developed in two ways: first, it is integrated across the membrane to arrive at an overall expression for ion flux, and second, it is differentiated to arrive at a second-order differential equation. In this way, the mathematical code (obtained after numerical discretization) becomes very robust and can easily be solved by any algebraic equation-solver.

\indent The integrated form of Eq. (\ref{eq:MS2}) becomes (after multiplying by $c_i$),

\begin{equation}
- \frac{c_{i,\text{d}}^\ominus-c_{i,\text{c}}^\ominus}{\delta_\text{m}} - \frac{z_i}{\delta_\text{m}} \: \int_0^{\Delta\phi_m} c_i \: \text{d}\phi = \frac{1}{D_{i-F}} \left( J_i - \langle c_i \rangle \: v_{F} \right) +  f_{i-m} \: J_i + \beta \cdot \left( \langle c_k \rangle \: J_i  - \langle c_i \rangle \: J_k \right)
\label{eq:MS3}
\end{equation}

\noindent where average concentrations are defined by Eq. (\ref{eq:average})a with $h$ replaced by $\delta_\text{m}$, and where $c_{i,\text{c/d}}^\ominus$ denotes the concentration of ion type i at the membrane/channel interface, but still just in the membrane, at the c- or d-side. After discretization, the integral involving d$\phi$ can be solved using the trapezoid rule. 

Differentiating Eq. (\ref{eq:MS2}) (multiplied by $c_i$) leads to

\begin{equation}
\frac{\partial^2 c_i }{\partial x^2} + z_i\: \frac{\partial}{\partial x} \left( c_i \frac{\partial{\phi}}{\partial{x}} \right)  = \left( \frac{v_F}{D_{i-F}}  + \beta \left( J_k  - J_i \right)\right) \frac{\partial c_i}{\partial x} 
\label{eq:MS5}
\end{equation}

\noindent where, based on Eq. (\ref{eq:LEN}), we made used of $\partial_x c_i=\partial_x c_k$ for a binary 1:1 salt and fixed membrane charge, $X$. Additionally, for the fluid we use an integrated version of Eq. (\ref{eq:fluid_2}), which leads to

\begin{equation}
-\frac{2}{\delta_\text{m}} \left(c_\text{d}^* - c_\text{c}^* \right) = -f_{F-m} \: v_F +  \sum_i \frac{1}{D_{i-F}}\left(\vphantom{\sum} J_i - v_F \langle c_i \rangle \right)
\label{eq:fluid_3}
\end{equation}

\noindent where $c_\text{c/d}^*$ is the salt concentration in either the c- or d-channel, right at the membrane/channel interface. In Eq. ({\ref{eq:fluid_3}}), we assume there is no hydrostatic pressure drop between the two channels. Thus, the pressure decay in {$y$}-direction through the spacer channel is assumed to be the same in each channel, to keep the difference between them (across the membrane) at zero. Interestingly, inside the membrane the hydrostatic pressure gradually changes from left to right, but to solve the membrane problem, we do not need to calculate it. We also do not need to solve a differentiated version of Eq. (\ref{eq:fluid_2}).

\indent Because of the symmetry of our geometry, the flux through one membrane of both ions together, $J_{\text{ions},m}$, equals the salt flux, $J_\text{salt}$, transported between the c- and d-channels. For ED, the ratio of $J_\text{salt}$ over $J_\text{ch}$ is the current efficiency, $\lambda$, while for RED, the inverse is defined as the salt flux efficiency, $\vartheta$~\citep{Tedesco2016}. Both properties can be analyzed as a local ($y$-dependent) efficiency, see Fig. \ref{fig:ED_single_pass},  but, to compare with data, they can also be defined as an average over the cell, after first averaging current and salt flux separately (see Eq. (18) in ref.~\citep{Tedesco2016}, and Fig.~\ref{fig:theta} for an example).

\subsection{Overall and boundary conditions}
\label{sec:BoundaryConditions}

\indent The cell pair voltage, $V_\text{CP}$, is the sum of all voltage differences across a cell pair and can be calculated as

\begin{equation}
	V_\text{CP} = 2 \; V_\text{T} \; \left( \Delta\phi_\text{c} + \Delta\phi_\text{D,c} + \Delta\phi_m - \Delta\phi_\text{D,d} + \Delta\phi_\text{d} \right)
	\label{eq:cp_voltage}
\end{equation}
\noindent where $\Delta\phi_\text{c}$, $\Delta\phi_m$ and $\Delta\phi_\text{d}$ are the potential drops over the concentrate half-channel, the inner coordinate of the membrane, and the diluate half-channel, respectively (evaluated as the difference between potential at a position ``right'' minus ``left''); $\Delta\phi_\text{D,c}$ and $\Delta\phi_\text{D,d}$ refer to the Donnan potentials arising at the two membrane/channel interfaces; the factor of two is because a cell pair is twice our computational domain. Though $V_\text{CP}$ is the same at each $y$-coordinate, how it splits out in its separate contributions, that depends on $y$-position.

\indent The Donnan potentials on each membrane/solution interface, $\Delta\phi_\text{D,c/d}$ are a function of the membrane charge density, $X$, and can be calculated as \cite{Helfferich1962,Biesheuvel2011b,Kontturi2008}

\begin{equation}
	\omega X = 2 c_\text{c/d}^* \sinh \left(\Delta\phi_\text{D,c/d}\right). 
	\label{eq:donnan}
\end{equation}
\indent The concentration of an ion i just in the membrane, $c_i^\ominus$, relates to that just outside according to Boltzmann's law, $c_{\text{c/d},i}^\ominus=c_\text{c/d}^* \: \exp\left(-z_i \Delta\phi_\text{D,{c/d}} \right)$. In Eq.~({\ref{eq:donnan}}) we assume validity of Boltzmann's law for ions as ideal point charges, and we do not include additional contributions to ion partitioning beyond the direct electrostatic effect.

\begin{figure}
\centering
\includegraphics[width=\textwidth]{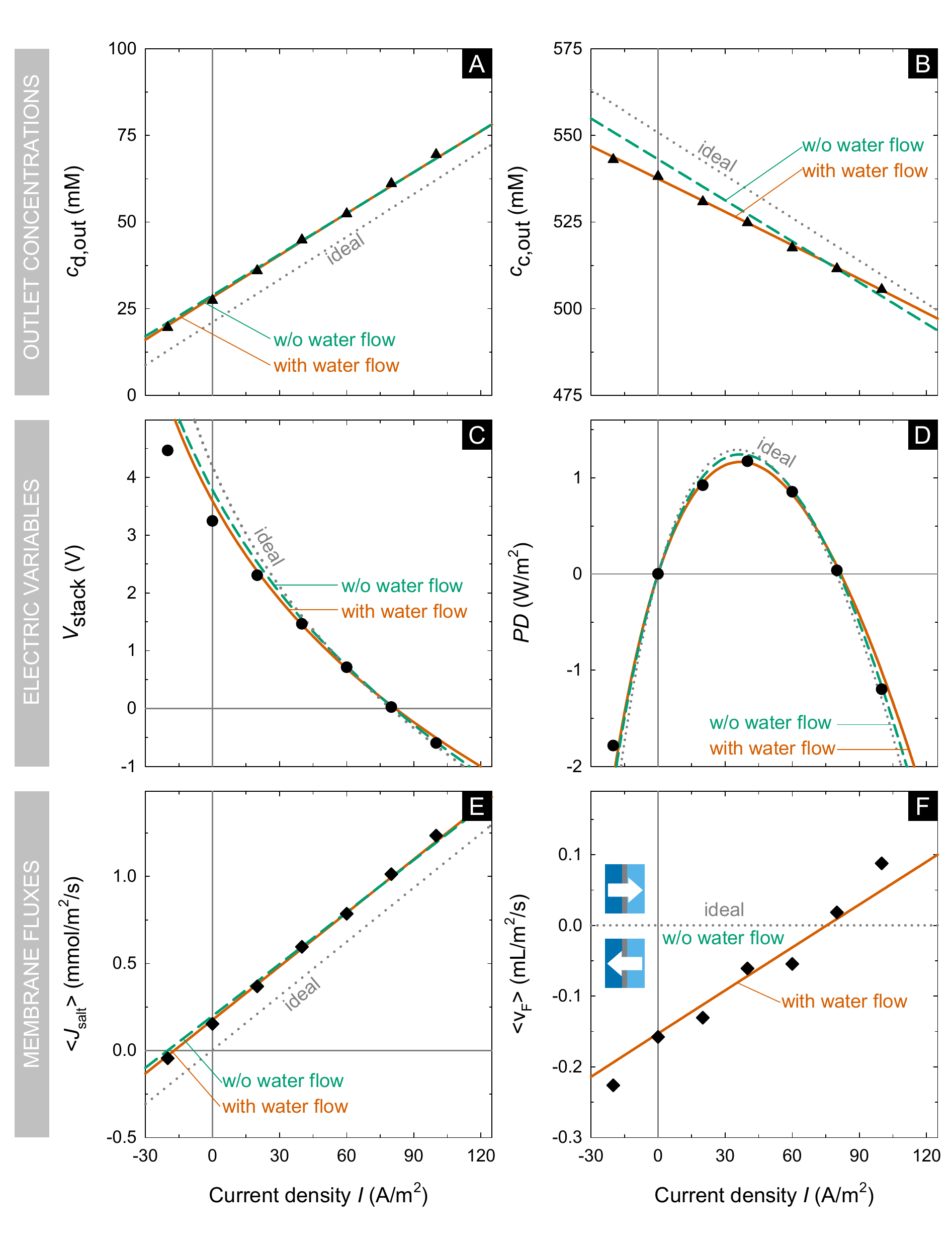}
\caption{Model validation for an RED stack with 25 cell pairs by comparison with data (symbols) of Veerman \textit{et al.} \citep{Veerman2009a} as function of average current density. A,B) River water and seawater outlet concentration, $c_\text{d,out}$, $c_\text{c,out}$. C) Stack voltage, $V_\text{stack}$. D) Power density, $PD$, per m$^2$ of total membrane area (both AEMs and CEMs combined). E) Average salt flux, $\langle J_\text{salt} \rangle $. F) Average membrane fluid velocity, $\langle v_{F} \rangle $. Lines are model predictions, where solid lines include coion and water transport; dashed lines: with coion transport, no water transport; dotted lines: no coion transport and no water transport, i.e., ideal membranes. The region with $\langle J_\text{ch} \rangle>0$ and $V_\text{stack}>0$ refers to RED-mode; $\langle J_\text{salt} \rangle <0$: ED-mode; the region with $V_\text{stack}<0$ is `forced RED', or `assisted RED'.}
\label{fig:FIG_validation_panel}
\end{figure}

\indent The above set of equations describes ion and water flow in a 2D model for the (R)ED cell and can be solved for any set of input conditions (flow rates, inlet salt concentrations, cell pair voltage $V_\text{CP}$ or average current $I$), together with values for ion diffusion coefficients, ion-membrane frictions, the fluid-membrane friction (water permeability), and membrane thickness and charge, $\delta_\text{m}$ and $X$. The differential equations are discretized in $x$- and $y$-direction, and the resulting set of algebraic equations solved. Like in ref.~\cite{Tedesco2016}, discretization in the $y$-direction is by the implicit (backward) Euler method, and in the $x$-direction we use a central difference method. Note that the cell pair voltage $V_\text{CP}$ is constant along the $y$-coordinate, as is expected for (R)ED cells with unsegmented electrodes. However, fluxes in $x$-direction (current, ion flux, fluid flow) vary with $y$-coordinate. Average current density, water velocity and salt flux, as presented in Figs.~\ref{fig:FIG_validation_panel} and \ref{fig:FIG_ED_time}, are values averaged from entrance to exit of the cell (over the entire $y$-coordinate).

\section{Results and discussion}
\label{sec:ResultsAndDiscussion}

\indent In the next section we show results of validation of our model against experimental data for RED and ED, as well as additional model calculations for RED at higher salt concentration, and for single-pass seawater desalination by ED. We address the influence of water and coion transport on process performance.  

\subsection{Model validation for Reverse Electrodialysis}
\label{sec:modelValidation}

\indent To validate the model, in which all fluxes through the membrane, including fluid flow, are self-consistently calculated from the applied driving forces (which all change between entrance and exit of the cell), calculation results are first of all compared with a representative data set for RED from Veerman \textit{et al.} \citep{Veerman2009a}. After having obtained a good fit of the full model to the data, we will reduce the model complexity and compare three cases: (1) the full model, which considers the fluxes of counterions, coions and water through the membrane; (2) a model where the flux of water (fluid flow) is set to zero; (3) a further simplified model, in which both the water flux and coion transport are set to zero, and coions are excluded from the membrane, thus simulating the case of ideal (100\% permselective) membranes. 

\indent The experiments by Veerman et al.~\citep{Veerman2009a} were done in a stack consisting of 25 cell pairs, containing Sefar Nitex spacers with a thickness of $\delta_\text{sp}=200$ $\mu$m and Fumasep FAD/FKD membranes with a thickness $\delta_m=80$ $\mu$m and membrane charge of $X\sim$4.0 M \citep{Guler2013}. Inlet salt concentrations are 21 mM and 551 mM of NaCl, for river and seawater. The membrane area is 10x10 cm$^2$ and the feed flow rate per channel 15.12 mL/min (without water flow through the membranes, the ``empty tube'' residence time is 7.94 s). The fluid flows from one corner of the spacer to its opposite end, with the main flow direction in the two channels at cross angles to one another. Thus, the flow profile is resembles cross-current rather than co-current. Still, in the present work we will assume co-current flow.

\indent For the membrane, the model has a total of six mobility factors, being the diffusion coefficients of counterion (ct) and coion (co) in the membrane ($D_{\text{ct}-F}$, $D_{\text{co}-F}$), the friction factor between counter- and coions, $\beta$, the friction of ions with the membrane matrix ($f_{\text{ct}-m}$, $f_{\text{co}-m}$), and the water-membrane friction (related to hydraulic permeability), $f_{F-m}$. Furthermore, as free parameters we have the membrane charge density $X$ and, for the spacer channel, the values of $D_\text{d}$ and $D_\text{e}$ (assumed the same for both ions). The model validation is performed by fitting five experimental conditions: two values of the diluate effluent concentration (at two different stack voltages), and two values of the concentrate effluent concentration, as well as the current when the stack is short-circuited (when $V_\text{stack}=0$). With these five constraints, the optimal parameters to fit the theory to the data are found using a Nelder-Mead method.

In the present work, we simplify the analysis by setting the ion-membrane friction coefficients to zero.  Thus in the model ions in the membrane have friction with the fluid and with other ions, but not with the membrane pore walls. For the ion-fluid diffusion coefficients in the membrane, the optimal values found are $D_\text{ct-F}=78$ $\mu$m$^2$/s and $D_\text{co-F}=162$ $\mu$m$^2$/s, which are respectively $\sim 5 \%$ and $\sim 10 \%$ of the diffusion coefficient at infinite dilution, $D_\infty$ (where $D_\infty=\sqrt{D_{\infty,\text{Na}^+} D_{\infty,\text{Cl}^-}}=1640$ $\mu$m$^2$/s). Thus, for the counterion, the diffusion coefficient in the membrane is reduced by a factor of $\sim$20, and for the coion by a factor of $\sim$10. Furthermore, we find the best fit using a membrane charge density of $X=4.2$ M, an ion-ion friction of $\beta \sim 0.60$ s$\cdot$m/$\mu$mol, and a water-membrane friction of $f_{F-m}=18$ Tmol$\cdot$s/m$^5$ (corresponding to a water permeability of $L_\text{p}=100$ mL/m$^2$/bar/hr). 

\indent For the spacer channel, optimal values are $D_\text{d}=1640$ $\mu$m$^2$/s and $D_\text{e}=515$ $\mu$m$^2$/s, where $D_\text{d}$ is larger than $D_\text{e}$ because it includes dispersion (which enhances concentration equalization). The ratio $D_\text{e}/D_\infty$ can be ascribed to a reduction in effective diffusion coefficient due to the spacer porosity $\epsilon$ and tortuosity factor $\tau$ \citep{Catalano2016}. In this case, we obtain for the spacer channel a value of $\epsilon/\tau \sim 0.314$. All of these values are very realistic, with the diffusion coefficients of the ion in the membrane, $D_{i-F}$ about 5-10\% of the value in free solution, as reported earlier \citep{Danielsson2009}. Membrane charge density is close to the value reported by G{\"u}ler \textit{et al.}~\citep{Guler2013}. The value for the water-membrane friction, $f_{F-m}$, will be discussed further on.

\begin{figure}
\centering
\includegraphics[width=0.7\textwidth]{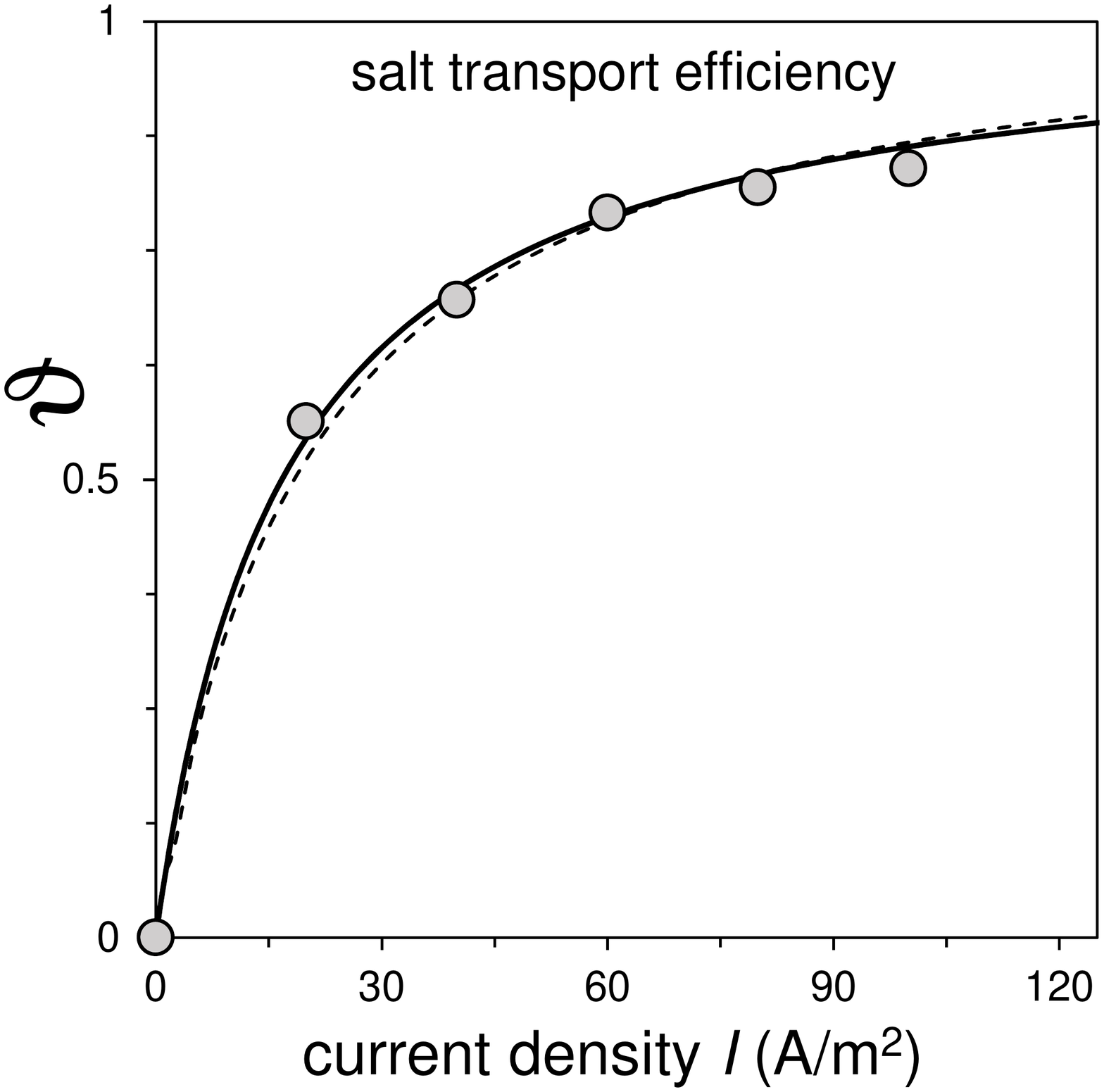}
\caption{Data and theory for the cell-averaged salt transport efficiency in reverse electrodialysis, $\vartheta$, as function of cell-averaged current $I$, based on data and theory in Fig.~\ref{fig:FIG_validation_panel}.}
\label{fig:theta}
\end{figure}

\indent Fig. \ref{fig:FIG_validation_panel} shows a comparison between model predictions and experimental data for the outlet salt concentrations, $c_\text{c,out}$ and $c_\text{d,out}$, stack voltage, $V_\text{stack}$, power density, $PD$, average salt flux, $\langle J_\text{salt} \rangle$, and fluid velocity through the membrane, $\langle v_F \rangle$. Model predictions agree well with data, not only in the RED regime, but also for the data in ED mode (where $\langle J_\text{salt} \rangle <0$), and in the `forced RED' regime ($V_\text{stack}<0$). Moreover, the model reproduces the non-linear dependence of voltage on current (Fig. \ref{fig:FIG_validation_panel}C) due to the change in internal resistance because of ion transport. In addition, Fig.~{\ref{fig:theta}} shows the very good fit of theory to data for salt flux efficiency, $\vartheta$, based on the results in Fig.~{\ref{fig:FIG_validation_panel}}. The salt flux efficiency, $\vartheta$, is about 70 \% at maximum power (for a current $\sim$40 A/m$^2$).

\subsection{Evaluation of ion and water velocities in RED}
\label{sec:velocities_RED}

\indent With the model validated against data, it is interesting to analyze in more detail microscopic properties of transport, such as ion and water velocities in the membranes. For this analysis, we chose the condition of maximum power production, i.e., at a current of 40 A/m$^2$, and a position half-way along the channel. Model calculations show that, across the membrane, the concentration of each ion decays quite linearly from the concentrate-side to the diluate-side, with a gradient of about 0.8 mM/$\mu$m. The electric potential also changes gradually, with a mere 5 mV drop across (the inner coordinate of) the membrane. The fluid velocity is about $v_F=$--0.075 $\mu$m/s (see Fig.~\ref{fig:FIG_validation_panel}F), directed towards the concentrate side. Both ions move in the other direction, towards the diluate side, with the counterion having a velocity of $v_\text{ct}\sim $0.12 $\mu$m/s. This velocity is almost independent of $x$-position in the membrane because the counterion concentration is quite constant across the membrane, and only slightly changes from $\sim$4.27 M to $\sim$4.20 M. However, for the coion the situation is totally different: it diffuses faster and its velocity is not constant. At the concentrate side, the coion has a velocity of $v_\text{co}=$1.3 $\mu$m/s, and near the diluate side it accelerates in the very last few percent of the membrane to a velocity almost 200 times higher. We do not know whether forces associated with this strong acceleration of the coion need to be considered in the transport theory, nor would we know how to do so. This coion acceleration is related to the linear decay of coion concentration, from around 65 mM on the concentrate-side, to $\sim$0.4 mM on the diluate side. This reduction in concentration, at constant ion flux (steady state), must lead to an increase in ion velocity that is inversely proportional to concentration. For the coion, despite its low average concentration, it is due to its high velocity ($\sim$2.5 $\mu$m/s in the middle of the membrane), that it nevertheless has a quite significant flux ($\sim$18\% of that of the counterion), i.e., the salt flux efficiency $\vartheta=J_\text{ch}/J_\text{salt}$ is only 70 \%~\citep{Tedesco2016}. 

\indent In our modeling framework it is possible to investigate the origin of the high flux of the coions in more detail. Focusing on a position half-way across the membrane, here the counterion concentration is 4234 mM and the coion concentration 34 mM (125$\times$ less than counterions). Three effect jointly explain the high coion flux (a fourth effect reduces the coion flux). I). The total force acting on a coion is around 9.2 times higher than for counterions. This can be explained as follows. The diffusional force on an ion scales with $F_\text{d}=-$d$ \ln{c} / $d$x=-1/c \cdot $d$ c / $d$x$ and while the gradient $-$d$c/$d$x$ is the same for counter- and coion, the prefactor $1/c$ is 125$\times$ higher for the coion, thus making the diffusional force higher for the coion by that factor. However, for counterions the electric field is an additional driving force which is about 11.4$\times$ larger. For coions, the electric field is in magnitude equally large, but relatively to diffusion, not 11.4$\times$ larger (as for counterions), but 11.4$\times$ smaller (and working against flow). II). Both ions move against the direction of fluid, which reduces their flux. While this effect is small for the coion (it goes between 20 and 2000$\times$ faster than the fluid), it is significant for the counterion, whose velocity relative to the membrane is around 61\% of its velocity relative to the fluid. III). Finally, the coion has a diffusion coefficient (inverse of ion-fluid friction coeficient) in the membrane that is 2.1$\times$ that of the counterion. IV). Whereas ion-ion friction hardly influences the counterion force balance, it consumes ~28\% of the force on coions (coions are retarded because of ion-ion friction with the much slower counterions). Combining these effects, we find that the coion flux is $\sim 9.2/125/0.61 \cdot 2.1 \cdot (1-0.28) = 18\%$ of the counterion flux.

\indent With the full model (that includes water flow and transport of both ions) successfully fitted to the data, it is now interesting to reduce the model complexity by first setting the water flow velocity to zero, and next also to exclude coion flow, and compare the three calculations. First, we set the water velocity through the membrane, $v_F$,  to zero in the model, to arrive at the dashed lines in Fig.~\ref{fig:FIG_validation_panel}. Without water flow, the two channels will have the same inflow as outflow volume flux, and thus the two lines for effluent concentration in Fig. \ref{fig:FIG_validation_panel}A--B become anti-parallel, i.e., their summation must be constant, equal to the sum of the two inlet concentrations. Notably, a significant change  of $c_{c,\text{out}}$ is predicted in this case (Fig. \ref{fig:FIG_validation_panel}B), and this difference is due to water flow through the membrane. The salt flux (Fig. \ref{fig:FIG_validation_panel}E) is hardly changed, thus also $\vartheta$ is almost the same.  At the condition of maximum power density, the current is slightly higher, as well as the power density (+7\%). 

\indent If coions are now also excluded from the membrane, power density further increases (another \mbox{+4 \%} compared to the ``no water flow''-case) and Fig.~\ref{fig:FIG_validation_panel}C shows that the cell voltage at zero current (i.e., the open circuit voltage, OCV) predicted by the full model is significantly lower than in the ideal case ($V_\text{stack}=3.6$ vs. $4.2$ V). 

\indent The results above demonstrate that the model can predict performance of an RED process in a wide range of applied currents, and that to describe data accurately, water flow is ideally included. Water flow reduces the power density (7\% at maximum power; and an additional 4\% because of coion leakage), and the water flow velocity $v_F$ in the membrane is not small: it is of the same order of magnitude as the counterion velocity, and has an effect on the predicted profiles for effluent concentration. 

\indent The procedure to fit the theory to the data resulted in a value for the water permeability of $f_{F-m}=18$ Tmol$\cdot$s/m$^5$. This number can be multiplied by the membrane thickness, $\delta_\text{m}$, and by $R_\text{g}T\sim$2500 J/mol, and then inverted, to give a water permeability of $L_\text{p}=100$ mL/m$^2$/hr/bar. This is almost 20x higher than what is reported in literature for commercial IEMs \citep{Xu2005}. At the end of this section we derive an effective pore size based on this value of $L_\text{p}$.

\subsection{Effect of water transport in RED with highly concentrated brines}

\begin{figure}[ht]
\centering
	\includegraphics[width=0.7\textwidth]{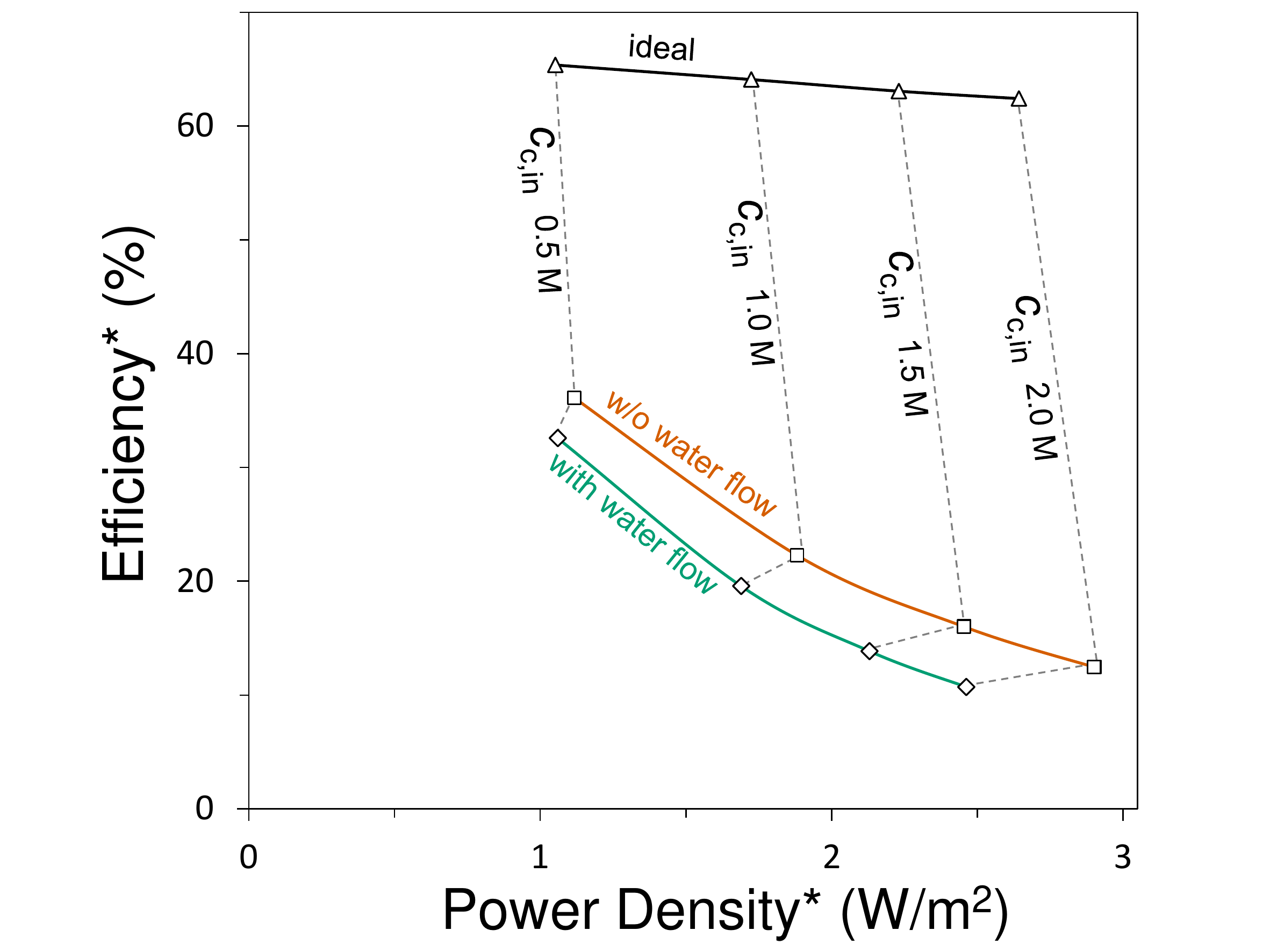}
\caption{Effect of inlet brine concentration, $c_\text{c,in}$, on the performance of the RED process. Efficiency is plotted versus power density, for conditions where the response product of efficiency times power is at a maximum.}
\label{fig:changing_c_in}
\end{figure}

\indent In this sub-section we analyze the importance of water flow and coion leakage for RED at a higher starting concentration of the concentrate stream (brine), up to 2.0 M. These high concentrations can be advantageous for energy production (assuming these highly concentrated brines are available)~\citep{Tedesco2016performance}. The exact analysis of these energy losses due to water and coion transport are of particular relevance for processes in which feed solutions are used in a closed loop circuit, for example in energy storage systems \citep{Kingsbury2015, vanEgmond2016}. Results are presented in Fig. \ref{fig:changing_c_in}, where, except for the brine concentration, all parameter settings are the same as in the previous section, including river water concentration \mbox{(21 mM)}.

\indent Instead of presenting full curves for stack voltage and power density versus current density, as in Fig.~\ref{fig:FIG_validation_panel}C--D, here we determine for each such curve the point where the response product (RP)~\citep{Veerman2009a} of power $\times$ efficiency is at a maximum. Efficiency and power at that point are presented in Fig. \ref{fig:changing_c_in}. Efficiency is defined as the electric power extracted, over the loss of Gibbs energy, which is based on a calculation of $2\: R_\text{g}\:T \:Q\: c\: \ln{c}$ over all inlet and outlet streams, and taking the difference (where $Q$ is the flow rate). Fig. \ref{fig:changing_c_in} shows that with increasing brine concentration, the power that can be extracted (at the point of maximum RP) increases, but efficiency goes down. For ideal membranes the efficiency is by far the highest, even though power (at the point of maximum RP) is similar to the cases with coion transport and with/without water flow. Comparing these two non-ideal cases, without water flow both efficiency and power density are higher than with water flow. Thus, for all salt concentrations, both coion and water transport play an important role, either bringing down process efficiency, power density, or both.

\subsection{Effect of water transport in single-pass ED for seawater desalination}
\label{sec:calculationsED_singlepass}

\begin{figure}
\centering
\includegraphics[width=\textwidth]{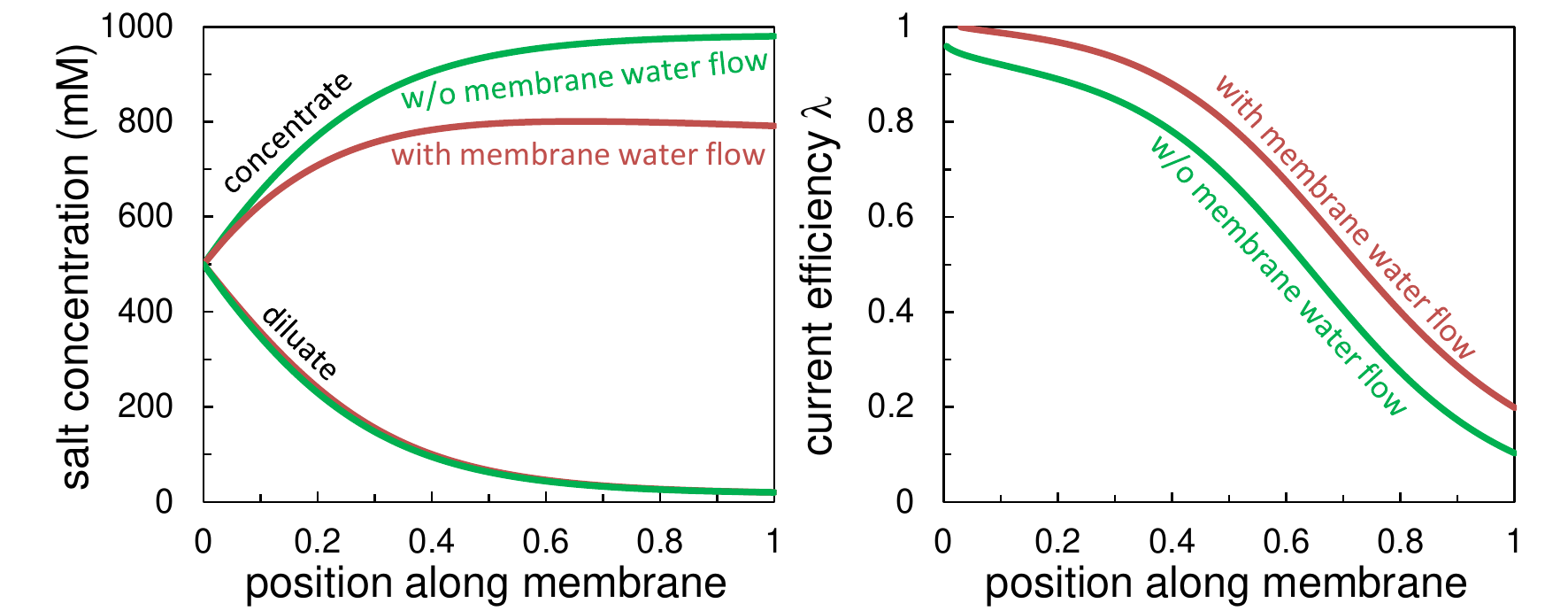}
\caption{Theoretical analysis of effect of membrane water transport on single pass seawater ED (to 20 mM in diluate channel). A) Profile of average salt concentration as function of position $y$. B) Local current efficiency $\lambda$.}
\label{fig:ED_single_pass}
\end{figure}

\indent In this section, we theoretically analyze a hypothetical steady-state single-pass ED experiment to obtain fresh water from seawater, both with and without water flow through the membrane, see \mbox{Fig. \ref{fig:ED_single_pass}}. 
The system is modeled using the same geometry and parameter settings obtained for the RED-calculations in the previous section. 

We first consider the situation that the concentration difference across the membrane is still small, at the same 40 A/m$^2$ current as analyzed in detail for RED. In this case, counterions move to the concentrate-side with a velocity of 0.1 $\mu$m/s, while water flows in the same direction, only by drag of the counterions (electro-osmosis), at a velocity of $\sim$0.7 $\mu$m/s. 
Since the water flows only because it is dragged by the ions (which are mainly counterions in the membrane) (there is no osmotic pressure difference across the membrane yet in this calculation), counterions must be faster than the water, though calculations show that the velocity difference is small. In this calculation, the only driving force for ions is the electric potential difference, which is just 0.63 mV across one membrane. Interestingly, the coion has a velocity relative to the water towards the diluate side, but relative to the membrane it moves towards the concentrate side, with a velocity about twice less than the counterion. This means that according to the calculation, current efficiency $\lambda$ is slightly larger than unity (1.01). For the counterion, the main frictional force is due to ion-water friction, but for the coion, ion-ion friction is about equally important as ion-water friction.

The situation just sketched, changes dramatically when we consider a situation of seawater desalination to freshwater at 20 mM. The (``empty tube'') residence time is set to 30 s (8 s in the previous RED-calculations), and the average current is $\sim$340 A/m$^2$ ($V_\text{CP}\sim$0.30 V), in order to reach a diluate concentration of 20 mM at the outlet of the cell. Because of the long residence time, and high current, water flow through the membrane is significant, such that the effluent flow rate in the diluate channel is only 75 \% of the inlet value. Analyzing exactly half-way in the cell, where the concentrate has an average concentration of 795 mM and the freshwater a concentration of 66 mM, we have a situation that in the membrane the counterion has a net velocity to the concentrate-side of $\sim$0.42 $\mu$m/s, but is outrun by the water, which flows at a velocity of $\sim$0.56 $\mu$m/s. Apparently, the concentration difference created earlier in the cell now pulls water through the membrane, in this way diluting the concentrate. At the same time, the coion moves to the freshwater side at a velocity which changes across the membrane from 1.1 $\mu$m/s on the concentrate side, to a value several hundred times higher on the diluate side. At this position, half-way in the cell, current efficiency $\lambda$ is 83 \%. Driving forces on the counterion (which relative to the water is moving to the freshwater side) are both diffusion and migration while for the coion it is only diffusion. 

\indent Without water flow, to reach 20 mM in the freshwater channel (with the same 30 s residence time), the average current must be higher, $\sim$380 A/m$^2$ ($V_\text{CP}\sim$0.41 V). Considering now a point halfway in the channel but for otherwise the same conditions, the counterion velocity is slightly higher, at $\sim$0.51 $\mu$m/s (towards the concentrate side), the water velocity is zero, and the coion has a velocity (towards the diluate side) that changes from $\sim$1.8 $\mu$m/s at the concentrate-side to a number 500$\times$ higher on the other side, close to 1 mm/s. Because the counterion cannot profit from the water flowing in the same direction, a much higher voltage drop across the (inner coordinate) of the membrane is needed, from 2.4 mV with water flow, to 15 mV in the case without water flow. Counterions move to the concentrate-side, driven by migration, while migration and diffusion both aid the coion to flow to the diluate side. Current efficiency is only $\lambda=71$ \%. With or without water flow, for the counterions only friction with the fluid is of significance (ion-ion friction perhaps 5\% relative to ion-fluid friction). However, for the coion, ion-ion friction is much more significant, up to 40 \% of the ion-fluid friction with water flow, and up to 50 \% without water flow.

\indent One unexpected observation of these calculations is the effect of water flow on energy consumption (current times voltage), and we calculated that without water flow, the electric energy input is 52\% higher. When calculated per m$^3$ of freshwater produced the difference is less, but still it is $\sim$16 \% higher. This might suggest that an ED process with water flow through the membrane makes desalination more energy-efficient, which seems a counterintuitive result. However, note that with water flow through the membrane the water recovery is reduced to 38 \% (from 50 \% without water flow). Desalination is less energy consuming at a lower water recovery and this explains why ED with membrane water flow seems to be more energy efficient. In a further calculation for the case with water flow, we increased the inlet diluate flowrate (and reduced the inlet concentrate flowrate) to reach at the outlet a water recovery of 50\% with a 20 mM final concentration. In this last case, the energy consumption is higher than in the case without membrane water flow, and the difference is 10 \% ($I \sim$453 A/m$^2$, $V_\text{CP}\sim$0.38 V, 66 \% of total inflow directed to diluate channel). Thus, according to this calculation, as long as the desalination objective function is correctly defined, water flow through membranes increases energy consumption.

\indent In conclusion, all of these calculations show that water transport has multiple effects on operation of an ED cell with high single-pass desalination. Further on we discuss results of a more common batch-wise seawater ED operation and compare with data. For the calculations just made we are not aware of suitable data for comparison and model validation. Especially of importance is the value of hydraulic permeability, $L_\text{p}$, and the question whether a lower value is not more realistic for commercial membranes. The calculations above show the dire consequences that water transfer can have, including a large dilution of the concentrate flow.

\subsection{Effect of water transport in batch-mode ED for seawater desalination}
\label{sec:calculationsED_batchmode}

\begin{figure}
\centering
	\includegraphics[width=\textwidth]{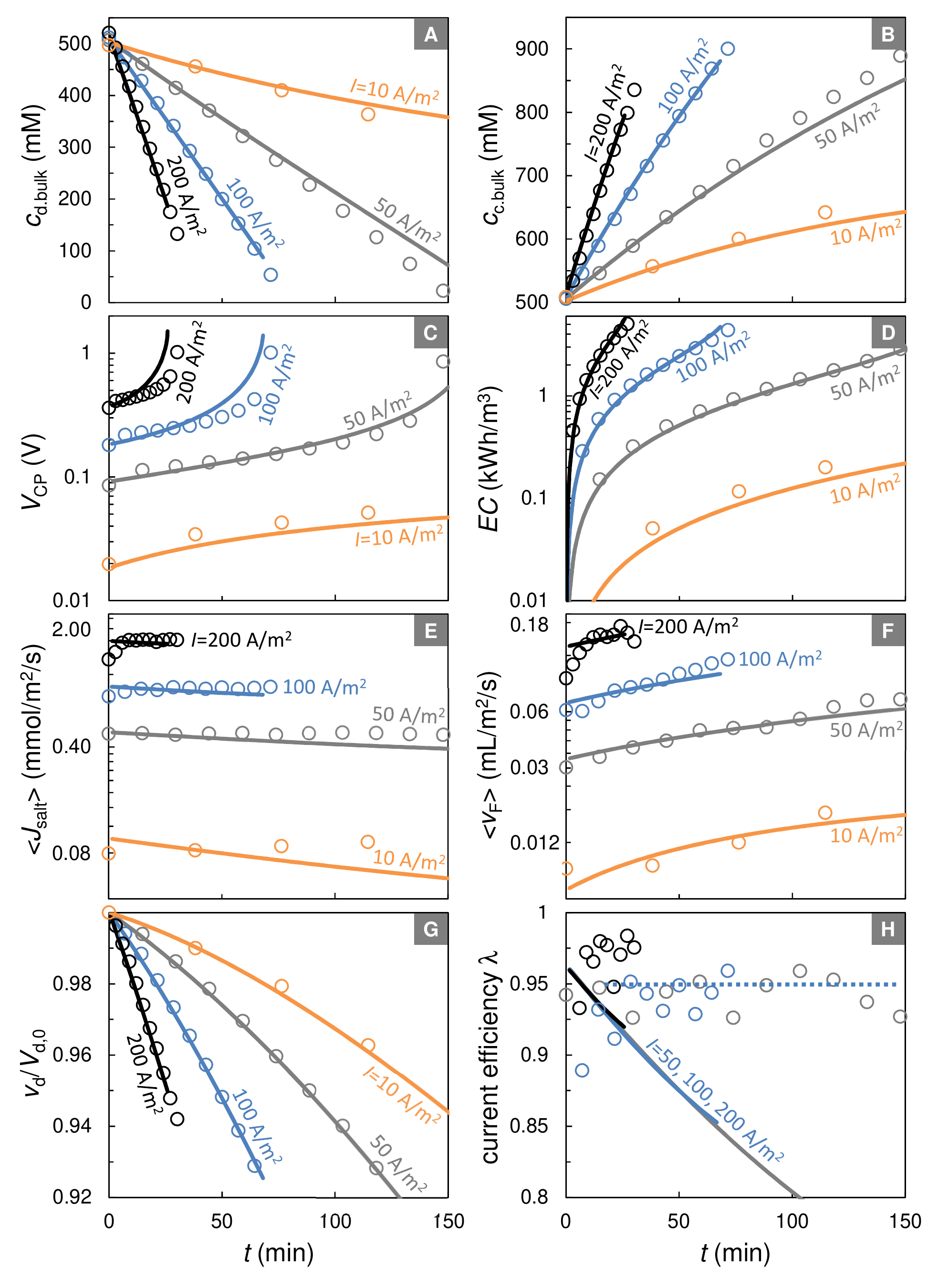}
\caption{Performance of ED process predicted by the full model including coion and water transport (full continuous lines), compared with data (symbols) of Galama \textit{et al.} \citep{Galama2014} as function of time of the experiment, $t$. A,B) Diluate and concentrate concentration, $c_\text{d,bulk}$, $c_\text{c,bulk}$. C) Cell pair voltage, $V_\text{CP}$. D) Energy consumption, $EC$, in kWh per m$^3$ of diluate. E) Average salt flux, $\langle J_\text{salt} \rangle$. F) Average membrane fluid velocity, $\langle v_\text{F} \rangle$. G) Normalized volume in the diluate tank, $V_{d}/V_{d,0}$. H) Current efficiency, $\lambda$. Dashed line refers to trendline of experimental data.}
\label{fig:FIG_ED_time}
\end{figure}

\indent The model calculations just discussed are for single pass seawater ED, for which no data are yet available. In existing experiment work, seawater is desalinated in batch mode~\citep{Galama2014}. Here the initial salt concentration in a diluate and a concentrate reservoir (``bulk'' in Fig. \ref{fig:FIG_ED_time}) is $c_\text{ini}\sim 510$ mM, and the effluent of the ED stack is returned to the reservoirs. The system is run at constant current, in the range of $J_\text{ch}=$10--200 A/m${^2}$. We make calculations with the same test conditions as reported by Galama \textit{et al.}~\citep{Galama2014}. The ED system consists of a 10-cell pair stack (membrane area 10x10 cm$^2$), equipped with Neosepta ACS/CMS membranes (Tokuyama Co., Japan; $\delta_m=130$ $\mu$m) and $\delta_\text{sp}=500$ $\mu$m spacers (Sefar Nitex 06-700/53). The stack is fed in batch mode from two 1 L reservoirs, with a constant feed flow rate of $Q_{in}=150$ mL/min (i.e., an inflow fluid velocity in the channel $v_\text{inflow}=0.5$ cm/s). The membrane thickness is $\delta_\text{m}=130$ $\mu$m, and $X \sim 5.0$ M. Theory includes an overall reservoir balance for volume and for total salt. For the ED stack, because of the low desalination per pass, we discretize using only one grid point in $y$-direction (i.e., conditions in the cell equal exit conditions). 

\indent Results of the fit of the model to the data are shown in Fig. \ref{fig:FIG_ED_time}. In particular, Fig. {\ref{fig:FIG_ED_time}}A--B show the time evolution of the salt concentration in the two reservoirs, $c_\text{c,bulk}$ and $c_\text{d,bulk}$, at different current densities. Clearly, high current densities (e.g., $I=$100, 200 A/m${^2}$) lead to fast desalination, but complete salt removal to obtain drinking water is not possible, due to the strong increase of the voltage, $V_\text{CP}$ (as shown in Fig. {\ref{fig:FIG_ED_time}}C). As a result, the energy consumption increases dramatically already during the first 60 min of the test (Fig. {\ref{fig:FIG_ED_time}}D). Note that the salt concentration in the reservoirs and the energy consumption are affected both by transport of salt and of water through the membranes. These effects are shown in Fig. {\ref{fig:FIG_ED_time}}E--F, where the salt flux, $J_\text{salt}$, and fluid velocity through the membrane, $v_{F}$, are reported. The final panels (Fig.  {\ref{fig:FIG_ED_time}}G--H) show the decrease in volume in the diluate reservoir, and the current efficiency (ratio of salt flux over current). Interestingly, current density of $I=100$ A/m${^2}$ (or higher) leads after 60 min experiment to a nearly $\sim$8 \% reduction of the volume of diluate (Fig. {\ref{fig:FIG_ED_time}}G), which constitutes a loss of target product in desalination processes. 

\indent A very good fit to the full data set can be obtained, though interestingly, according to the theory the current efficiency $\lambda$ decreases in time (solid lines in Fig. {\ref{fig:FIG_ED_time}}H), while the data show that the current efficiency is rather constant, at around 95 \% (dashed lines, Fig. {\ref{fig:FIG_ED_time}}H). The fit of theory to data results in the following parameter settings, for the membrane: $D_\text{ct-F}=50$ $\mu$m$^2$/s, $D_\text{co-F}=75$ $\mu$m$^2$/s, $X=5.28$ M, $\beta=0.60$ s$\cdot$m/$\mu$mol and $L_\text{p}=4.7$ mL/m$^2$/bar/hr ($f_{F-m}=234$ Tmol$\cdot$s/m$^5$), and for the spacer: $D_\text{d}=525$ $\mu$m$^2$/s and $D_\text{e}=178$ $\mu$m$^2$/s. Interestingly, the derived value for water permeability $L_\text{p}$ is now in line with values  of commercial membranes \citep{Xu2005}, unlike the value of $L_\text{p}$ used to fit the RED-experiments. The values of the diffusion and electromigration coefficients in the channel, $D_\text{d}$ and $D_\text{e}$, are about three times lower than in the RED-experiments. In the membrane, $D_\text{co-F}$ and $D_\text{ct-F}$ are slightly lower than for the RED-experiment, and again we find $D_\text{co-F}>D_\text{ct-F}$. The ion-ion friction in the membrane, $\beta$, is the same as before. The fitted value for $X$ is higher than for the RED-experiments, and is in line with the higher $X$ reported for Neosepta membranes relative to Fumasep membranes.

\indent Finally, we use the Hagen-Poiseuille equation for cylindrical straight pores and calculate the equivalent pore size $d_\text{eq}$ for the derived value of $L_\text{p}$. In all cases we assume a membrane porosity $\epsilon_m$ of 30~\%. Using $d_\text{eq}^2=32\cdot L_\text{p,exp}\cdot \delta_m\cdot \mu_w / \epsilon_m$ where $\mu_w$ is the dynamic viscosity of water ($\sim$1 mPa.s), we arrive for the two membranes at an equivalent diameter of $d_\text{eq}\sim$1.5 nm (Fumasep, RED-experiment) and 0.43 nm (Neosepta, ED-experiment). These are realistic numbers for the pore size of (R)ED membranes, considering also that due to pore tortuosity (not included in this calculation) the actual pore size will be larger than $d_\text{eq}$.

\clearpage

\section{Conclusions}
\label{sec:Conclusions}

In the present work we set up a fundamental description of transport phenomena in the (R)ED process, and investigated the effect of water transport through ion-exchange membranes on process performance. In the case of negligible ion-ion friction, we showed how the Maxwell-Stefan (MS) approach is formally equivalent to the hydrodynamic theory for hindered transport~\citep{Deen1987}. The present model self-consistently describes fluid flow as function of the driving forces on the water, being osmotic and hydrostatic pressure gradients, and as function of the frictions with the membrane matrix and with ions. We showed how, despite the extremely low water permeability of IEMs, the velocity of water in the membrane is not negligible, and has the same order of magnitude of the velocity of counterions. The model shows a reasonable agreement with two separate data sets for ED and RED, though to fit the RED-data the hydraulic permeability was possibly chosen too high. Direct ion-membrane friction was left out of the calculation, as well as non-electrostatic contributions to the partition coefficient at the membrane-solution interface, though both effects are probably of importance. We also assumed absence of a hydrostatic pressure difference across the membrane, and assumed perfect co-current flow conditions. Spacer shadow effects, which can reduce the spacer/membrane interfacial area, should also be considered in future work. 
To further validate the model, an extended data set using the same membranes and setup, covering both ED and RED modes, is required.

\section*{Acknowledgments}

This work was performed in the cooperation framework of Wetsus, European Centre of Excellence for Sustainable Water Technology (www.wetsus.eu). Wetsus is co--funded by the Dutch Ministry of Economic Affairs and Ministry of Infrastructure and Environment, the Province of Frysl\^{a}n, and the Northern Netherlands Provinces. The authors thank the participants of the research theme ``Blue Energy'' for fruitful discussions and financial support.


\bibliographystyle{elsarticle-num} 
\bibliography{library}

\begin{thebibliography}{10}
\expandafter\ifx\csname url\endcsname\relax
  \def\url#1{\texttt{#1}}\fi
\expandafter\ifx\csname urlprefix\endcsname\relax\def\urlprefix{URL }\fi
\expandafter\ifx\csname href\endcsname\relax
  \def\href#1#2{#2} \def\path#1{#1}\fi

\bibitem{Tedesco2016}
M.~Tedesco, H.~V.~M. Hamelers, P.~M. Biesheuvel, {Nernst-Planck transport
  theory for (reverse) electrodialysis: I. Effect of co-ion transport through
  the membranes}, Journal of Membrane Science 510 (2016) 370--381.

\bibitem{Sonin1968}
A.~A. Sonin, R.~F. Probstein, {A hydrodynamic theory of desalination by
  electrodialysis}, Desalination 5 (1968) 293--329.

\bibitem{evans2006}
C.~E. Evans, R.~D. Noble, S.~Nazeri-Thompson, B.~Nazeri, C.~A. Koval, Role of
  conditioning on water uptake and hydraulic permeability of
  nafion{\textregistered} membranes, Journal of Membrane Science 279 (2006)
  521--528.

\bibitem{Izquierdo2012}
M.~A. Izquierdo-Gil, V.~M. Barrag{\'{a}}n, J.~P.~G. Villaluenga, M.~P. Godino,
  {Water uptake and salt transport through Nafion cation-exchange membranes
  with different thicknesses}, Chemical Engineering Science 72 (2012) 1--9.

\bibitem{han2015}
L.~Han, S.~Galier, H.~Roux-de Balmann, Ion hydration number and electro-osmosis
  during electrodialysis of mixed salt solution, Desalination 373 (2015)
  38--46.

\bibitem{Kedem1961}
O.~Kedem, A.~Katchalsky, {A physical interpretation of the phenomenological
  coefficients of membrane permeability.}, The Journal of General Physiology 45
  (1961) 143--179.

\bibitem{Holt1981}
T.~Holt, S.~K. Ratkje, K.~S. F{\o}rland, T.~{\O}stvold, Hydrostatic pressure
  gradients in ion exchange membranes during mass and charge transfer, Journal
  of Membrane Science 9 (1981) 69--82.

\bibitem{jiang2015}
C.~Jiang, Q.~Wang, Y.~Li, Y.~Wang, T.~Xu, Water electro-transport with hydrated
  cations in electrodialysis, Desalination 365 (2015) 204--212.

\bibitem{Peters2016}
P.~B. Peters, R.~van Roij, M.~Z. Bazant, P.~M. Biesheuvel, Analysis of
  electrolyte transport through charged nanopores, Phys. Rev. E 93 (2016)
  053108.

\bibitem{Tanaka2004pressure}
Y.~Tanaka, {Pressure distribution, hydrodynamics, mass transport and solution
  leakage in an ion-exchange membrane electrodialyzer}, Journal of Membrane
  Science 234 (2004) 23--39.

\bibitem{Kodym2012}
R.~Kod{\'{y}}m, P.~P{\'{a}}nek, D.~{\v{S}}nita, D.~Tvrzn{\'{i}}k, K.~Bouzek,
  {Macrohomogeneous approach to a two-dimensional mathematical model of an
  industrial-scale electrodialysis unit}, Journal of Applied Electrochemistry
  42 (2012) 645--666.

\bibitem{Veerman2008parasitic}
J.~Veerman, J.~W. Post, M.~Saakes, S.~J. Metz, G.~J. Harmsen, {Reducing power
  losses caused by ionic shortcut currents in reverse electrodialysis stacks by
  a validated model}, Journal of Membrane Science 310 (2008) 418--430.

\bibitem{Teorell1953}
T.~Teorell, {Transport processes in ionic membranes}, Progress in Biophysics
  and Biophysical Chemistry 3 (1953) 305--369.

\bibitem{Probstein1972}
R.~F. Probstein, A.~A. Sonin, E.~Gur-Arie, A turbulent flow theory of
  electrodialysis, Desalination 11 (1972) 165--187.

\bibitem{Lacey1980}
R.~E. Lacey, {Energy by reverse electrodialysis}, Ocean Engineering 7 (1980)
  1--47.

\bibitem{Pintauro1984}
P.~N. Pintauro, D.~N. Bennion, {Mass transport of electrolytes in membranes. 1.
  Development of mathematical transport model}, Industrial and Engineering
  Chemistry Fundamentals 23 (1984) 230--234.

\bibitem{Guzman-Garcia1990}
A.~G. Guzman-Garcia, P.~N. Pintauro, M.~W. Verbrugge, R.~F. Hill, {Development
  of a space-charge transport model for ion-exchange membranes}, AIChE Journal
  36 (1990) 1061--1074.

\bibitem{Higa1990}
M.~Higa, A.~Tanioka, K.~Miyasaka, {A study of ion permeation across a charged
  membrane in multicomponent ion systems as a function of membrane charge
  density}, Journal of Membrane Science 49 (1990) 145--169.

\bibitem{Fila2003}
V.~F\'{\i}la, K.~Bouzek, {A mathematical model of multiple ion transport across
  an ion-selective membrane under current load conditions}, Journal of Applied
  Electrochemistry 33 (2003) 675--684.

\bibitem{Volgin2005}
V.~M. Volgin, A.~D. Davydov, {Ionic transport through ion-exchange and bipolar
  membranes}, Journal of Membrane Science 259 (2005) 110--121.

\bibitem{Fidaleo2005}
M.~Fidaleo, M.~Moresi, {Optimal strategy to model the electrodialytic recovery
  of a strong electrolyte}, Journal of Membrane Science 260 (2005) 90--111.

\bibitem{Veerman2011}
J.~Veerman, M.~Saakes, S.~J. Metz, G.~J. Harmsen, {Reverse electrodialysis: A
  validated process model for design and optimization}, Chemical Engineering
  Journal 166 (2011) 256--268.

\bibitem{Tanaka2015}
Y.~Tanaka, M.~Reig, S.~Casas, C.~Aladjem, J.~L. Cortina, {Computer simulation
  of ion-exchange membrane electrodialysis for salt concentration and reduction
  of RO discharged brine for salt production and marine environment
  conservation}, Desalination 367 (2015) 76--89.

\bibitem{mehta1976}
G.~D. Mehta, T.~F. Morse, E.~A. Mason, M.~H. Daneshpajooh, {Generalized
  Nernst--Planck and Stefan--Maxwell equations for membrane transport}, The
  Journal of Chemical Physics 64 (1976) 3917--3923.

\bibitem{Meares1981}
P.~Meares, Coupling of ion and water fluxes in synthetic membranes, Journal of
  Membrane Science 8 (1981) 295--307.

\bibitem{Kraaijeveld1995}
G.~Kraaijeveld, V.~Sumberova, S.~Kuindersma, H.~Wesselingh, {Modelling
  electrodialysis using the Maxwell-Stefan description}, The Chemical
  Engineering Journal and the Biochemical Engineering Journal 57 (1995)
  163--176.

\bibitem{Amundson2003}
N.~R. Amundson, T.-W. Pan, V.~I. Paulsen, {Diffusing with Stefan and Maxwell},
  AIChE Journal 49 (2003) 813--830.

\bibitem{delacourt2008}
C.~Delacourt, J.~Newman, Mathematical modeling of a cation-exchange membrane
  containing two cations, Journal of the Electrochemical Society 155 (2008)
  B1210--B1217.

\bibitem{Galama2013}
A.~H. Galama, J.~W. Post, M.~A. Cohen~Stuart, P.~M. Biesheuvel, {Validity of
  the Boltzmann equation to describe Donnan equilibrium at the
  membrane-solution interface}, Journal of Membrane Science 442 (2013)
  131--139.

\bibitem{Thibault2015}
K.~Thibault, H.~Zhu, A.~Szymczyk, G.~Li, The averaged potential gradient
  approach to model the rejection of electrolyte solutions using
  nanofiltration: Model development and assessment for highly concentrated feed
  solutions, Separation and Purification Technology 153 (2015) 126--137.

\bibitem{Deen1987}
W.~M. Deen, Hindered transport of large molecules in liquid-filled pores, AIChE
  Journal 33 (1987) 1409--1425.

\bibitem{Brenner1977}
H.~Brenner, L.~J. Gaydos, The constrained brownian movement of spherical
  particles in cylindrical pores of comparable radius. models of the diffusive
  and convective transport of solute molecules in membranes and porous media,
  Journal of Colloid and Interface Science 58 (1977) 312--356.

\bibitem{Noordman2002}
T.~R. Noordman, J.~A. Wesselingh, {Transport of large molecules through
  membranes with narrow pores: The Maxwell-Stefan description combined with
  hydrodynamic theory}, Journal of Membrane Science 210 (2002) 227--243.

\bibitem{Lanteri2009}
Y.~Lanteri, P.~Fievet, A.~Szymczyk, Evaluation of the steric, electric, and
  dielectric exclusion model on the basis of salt rejection rate and membrane
  potential measurements, Journal of Colloid and Interface Science 331 (2009)
  148--155.

\bibitem{Szymczyk2010}
A.~Szymczyk, H.~Zhu, B.~Balannec, Ion rejection properties of nanopores with
  bipolar fixed charge distributions, The Journal of Physical Chemistry B 114
  (2010) 10143--10150.

\bibitem{Biesheuvel2011b}
P.~M. Biesheuvel, {Two-fluid model for the simultaneous flow of colloids and
  fluids in porous media.}, Journal of Colloid and Interface Science 355 (2011)
  389--95.

\bibitem{Yaroshchuk2011}
A.~E. Yaroshchuk, Transport properties of long straight nano-channels in
  electrolyte solutions: A systematic approach, Advances in Colloid and
  Interface Science 168 (2011) 278--291.

\bibitem{Sonin1976selegny}
{A. A. Sonin in: E. S{\'e}l{\'e}gny}, {Charged Gels and Membranes I}, D. Reidel
  Publishing Company, 1976.

\bibitem{Spruijt2014}
E.~Spruijt, P.~M. Biesheuvel, Sedimentation dynamics and equilibrium profiles
  in multicomponent mixtures of colloidal particles, Journal of Physics:
  Condensed Matter 26 (2014) 075101.

\bibitem{Schlogl1964}
R.~Schl{\"o}gl, {Stofftransport durch Membranen}, Dr. Dietrich Steinkopff
  Verlag.

\bibitem{Helfferich1962}
F.~G. Helfferich, {Ion Exchange}, Mc Graw-Hill, London, 1962.

\bibitem{Kontturi2008}
K.~Kontturi, L.~Murtom\"{a}ki, J.~A. Manzanares, {Ionic Transport Processes: In
  Electrochemistry and Membrane Science}, OUP Oxford, 2008.

\bibitem{Veerman2009a}
J.~Veerman, R.~M. de~Jong, M.~Saakes, S.~J. Metz, G.~J. Harmsen, {Reverse
  electrodialysis: Comparison of six commercial membrane pairs on the
  thermodynamic efficiency and power density}, Journal of Membrane Science 343
  (2009) 7--15.

\bibitem{Guler2013}
E.~G{\"u}ler, R.~Elizen, D.~A. Vermaas, M.~Saakes, K.~Nijmeijer,
  Performance-determining membrane properties in reverse electrodialysis,
  Journal of membrane Science 446 (2013) 266--276.

\bibitem{Catalano2016}
J.~Catalano, H.~V.~M. Hamelers, A.~Bentien, P.~M. Biesheuvel, {Revisiting
  Morrison and Osterle 1965: the efficiency of membrane-based electrokinetic
  energy conversion}, Journal of Physics: Condensed Matter 28 (2016) 324001.

\bibitem{Danielsson2009}
C.-O. Danielsson, A.~Dahlkild, A.~Velin, M.~Behm, A model for the enhanced
  water dissociation on monopolar membranes, Electrochimica Acta 54~(11) (2009)
  2983 -- 2991.

\bibitem{Xu2005}
T.~Xu, {Ion exchange membranes: State of their development and perspective},
  Journal of Membrane Science 263 (2005) 1--29.

\bibitem{Tedesco2016performance}
M.~Tedesco, C.~Scalici, D.~Vaccari, A.~Cipollina, A.~Tamburini, G.~Micale,
  {Performance of the first Reverse Electrodialysis pilot plant for power
  production from saline waters and concentrated brines}, Journal of Membrane
  Science 500 (2016) 33--45.

\bibitem{Kingsbury2015}
R.~S. Kingsbury, K.~Chu, O.~Coronell, {Energy storage by reversible
  electrodialysis: The concentration battery}, Journal of Membrane Science 495
  (2015) 502--516.

\bibitem{vanEgmond2016}
J.~W. van Egmond, M.~Saakes, S.~Porada, T.~Meuwissen, C.~J.~N. Buisman,
  H.~V.~M. Hamelers, {The concentration gradient flow battery as electricity
  storage system: Technology potential and energy dissipation}, Journal of
  Power Sources 325 (2016) 129--139.

\bibitem{Galama2014}
A.~H. Galama, M.~Saakes, H.~Bruning, H.~H.~M. Rijnaarts, J.~W. Post, Seawater
  predesalination with electrodialysis, Desalination 342 (2014) 61--69.

\end{thebibliography}

\end{document}